\documentclass[aps,pra,twocolumn,superscriptaddress,nofootinbib]{revtex4-2}

\usepackage{graphicx}	
\usepackage{amssymb}	
\usepackage{amsfonts}	
\usepackage{amsmath}	
\usepackage{xcolor}		
\usepackage[T1]{fontenc}	
\usepackage[utf8]{inputenc} 
\usepackage{makecell} 
\usepackage{booktabs} 

\usepackage[english]{babel} 	
\usepackage[colorlinks=true,linkcolor=blue,citecolor=blue]{hyperref} 

\usepackage[caption=false]{subfig}		
\usepackage{braket}		
\usepackage{bm}		
\usepackage[load-configurations = abbreviations]{siunitx}	
\begin{document}

\title{Breaking Rayleigh’s law with spatially correlated disorder to control phonon transport}

\author{S. Th\'ebaud}
\email[E-mail: ]{simon.thebaud@insa-rennes.fr}
\affiliation{Materials Science and Technology Division, Oak Ridge National Laboratory, Oak Ridge, Tennessee 37831, USA}
\affiliation{INSA Rennes, Institut Foton, UMR 6082, 35700 Rennes, France}
\author{L. Lindsay}
\affiliation{Materials Science and Technology Division, Oak Ridge National Laboratory, Oak Ridge, Tennessee 37831, USA}
\author{T. Berlijn}
\email[E-mail: ]{berlijnt@ornl.gov}
\affiliation{Center for Nanophase Materials Sciences, Oak Ridge National Laboratory, Oak Ridge, Tennessee 37831, USA}
\affiliation{Computational Sciences and Engineering Division, Oak Ridge National Laboratory, Oak Ridge, Tennessee 37831, USA}

\begin{abstract}
Controlling thermal transport in insulators and semiconductors is crucial for many technological fields such as thermoelectrics and thermal insulation, for which a low thermal conductivity ($\kappa$) is desirable. A major obstacle for realizing low $\kappa$ materials is Rayleigh’s law, which implies that acoustic phonons, which carry most of the heat, are insensitive to scattering by point defects at low energy. We demonstrate, with large scale simulations on tens of millions of atoms, that isotropic long-range spatial correlations in the defect distribution can dramatically reduce phonon lifetimes of important low-frequency heat-carrying modes, leading to a large reduction of $\kappa$ – potentially an order of magnitude at room temperature. We propose a general and quantitative framework for controlling thermal transport in complex functional materials through structural spatial correlations, and we establish the optimal functional form of spatial correlations that minimize $\kappa$. We end by briefly discussing experimental realizations of various correlated structures. 
\end{abstract}

\pacs{}
\maketitle

\textit{Introduction.} Thermal transport properties of solids are crucially important to a range of technologies. For some applications, such as electronics or fusion reactors, high thermal conductivities are desirable to evacuate heat from its source \cite{mooreEmergingChallengesMaterials2014,huThermalConductivityTungsten2017,wangNumericalSimulationsThermal2021}. For others, like thermal barriers and thermoelectric modules, high thermal resistivity is critical \cite{snyderComplexThermoelectricMaterials2008,beekmanInorganicCrystalsGlass2017}. In the latter cases, relevant materials are generally electronic insulators and semiconductors and thermal transport is dominated by lattice vibrations (phonons). Defects and disorder are often engineered in such materials to scatter heat-carrying phonons and reduce the lattice thermal conductivity \cite{tanHighThermoelectricPerformance2018,kleinkeNewBulkMaterials2010,vineisNanostructuredThermoelectricsBig2010}.

\begin{figure}
\centering
\includegraphics[width=0.6\columnwidth]{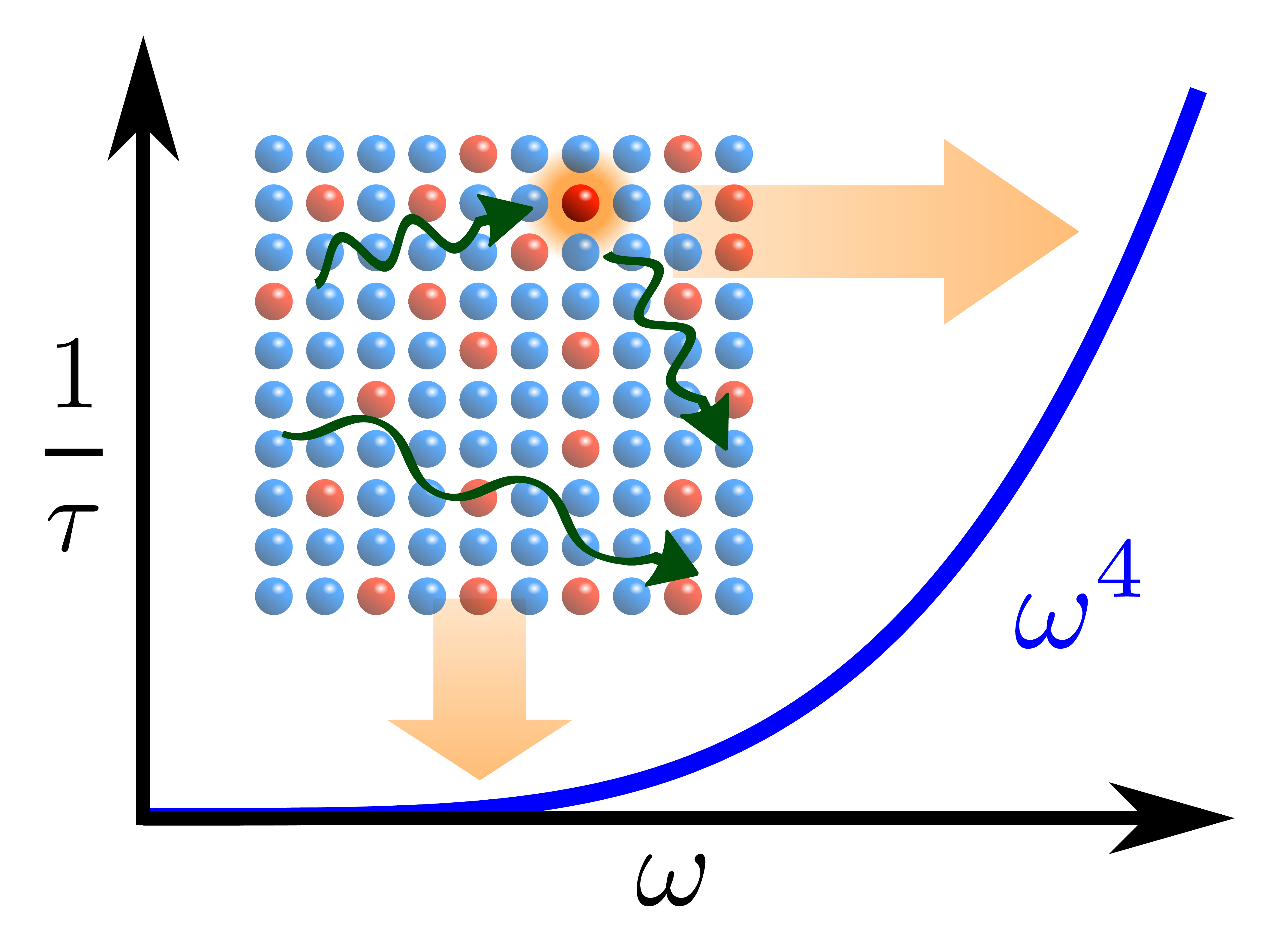}
\caption{The $\omega^4$ Rayleigh's law for phonon-disorder inverse lifetimes. Inset: low-$\omega$ phonons are insensitive to randomly distributed point defects, contrary to high-$\omega$ phonons.}
\label{fig1}
\end{figure}

However, a major impediment to achieving ultralow conductivities is Rayleigh’s law which, in this context, states that phonons are scattered by point defects at a rate proportional to the fourth power of their frequency: $1/\tau \propto \omega^4$ \cite{klemensScatteringLowFrequencyLattice1955}. This law is most famous in the context of light propagation, as it is responsible for the blue color of the sky. In solids, it is widely held to hold true for low frequency phonons scattering from atomic-scale defects such as vacancies or substitutional atoms. Because $\omega^4$ decreases rapidly with $\omega$, Rayleigh’s law implies that such disorder is inefficient at scattering acoustic phonons -- which typically carry most of the heat -- at low energy (see Fig.~\ref{fig1}) \cite{thebaudPerturbationTheoryThermal2022}. Consequently, researchers have sought to lower the lifetimes of these long-wavelength phonons by introducing disordered nanostructures of various size such as granular structures, dislocations, or nanoparticles \cite{chenManipulationPhononTransport2018,poudelHighThermoelectricPerformanceNanostructured2008,joshiEnhancedThermoelectricFigureofMerit2008,kimDenseDislocationArrays2015,zhengDefectEngineeringThermoelectric2021}. Based largely on the rule of thumb that phonons are most efficiently scattered by defects of size comparable to their wavelength, attempts have been made at combining these different types of disorder to target acoustic phonons over multiple frequency ranges, with various degrees of success. Thus far, experimental efforts have received little theoretical guidance, instead relying on empirical models and trial and error synthesis and characterization \cite{dresselhausNewDirectionsLowDimensional2007,zebarjadiPerspectivesThermoelectricsFundamentals2012,minnichBulkNanostructuredThermoelectric2009}.

\begin{figure*}
\centering
\includegraphics[width=1.8\columnwidth]{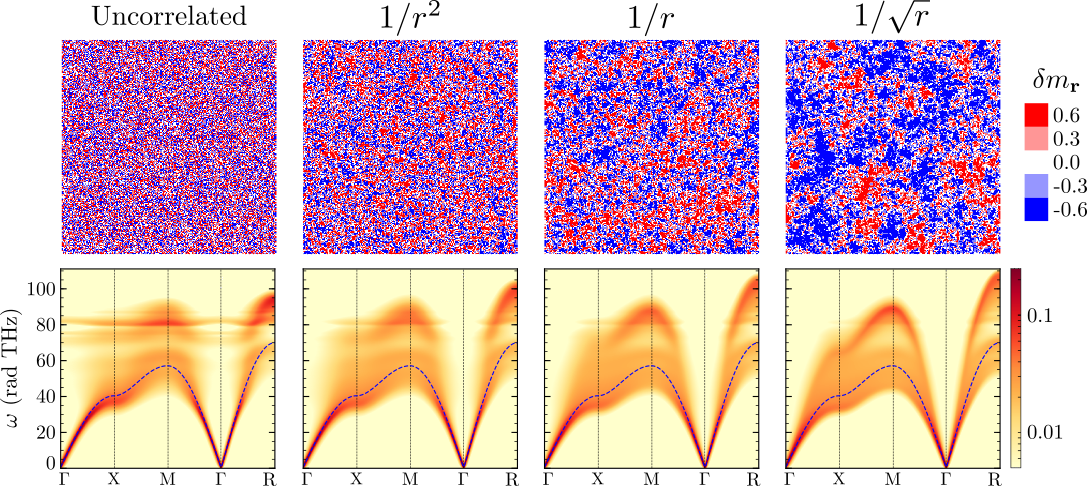}
\caption{Top: 2D cuts of the mass distribution in supercells of size $200 \times 200 \times 200$ atoms (each pixel representing an atom) with several power-law decaying correlations $C(\bm{r}) \propto 1/r^\alpha$ with $\alpha = \infty$ (uncorrelated) and $\alpha = 2$, $1$, and $0.5$. The color scale denotes the discrete relative difference from the average mass, with five atomic species present in the supercell. Bottom: the phonon spectral function $A(\bm{k},\omega)$ calculated by the CPGF method, together with the VCA dispersion (dashed blue line).}
\label{fig2}
\end{figure*}

As is well-known in optics, spatially correlated media can be used to manipulate wave scattering and propagation \cite{vynckLightCorrelatedDisordered2021,dikopoltsevLocalizationVirtualTransitions2019}. In the context of electron and phonon transport, some perturbative field theoretical calculations have been carried out for short-range and power-law decaying correlations in the 1980's, but the main focus was on Anderson localization properties, not on the reduction of electrical or thermal conductivities \cite{johnWavePropagationLocalization1983,chuEffectCorrelationsLocalization1989}. More recently, work has been devoted to studying phonon transport in one-dimensional chains with various forms of correlated disorder \cite{herrera-gonzalezControllingSizeScaling2020,savinHeatConductionDiatomic2017,albuquerqueEnhancedLocalizationEnergy2015,izrailevAnomalousTransportLowdimensional2005}. However, this field also has a strong fundamental focus since one-dimensional or quasi-one-dimensional systems are difficult to synthesize, let alone scale up industrially. Most technological applications require three-dimensional bulk compounds that typically display relatively isotropic properties. In particular, the formalism of spatial correlations has not been brought to bear on the long-standing effort to impede heat conduction in bulk semiconductors and insulators.

In this work, we show that Rayleigh’s law can be broken at all frequencies by atomic mass point defects with isotropic long-range spatial correlations, potentially yielding a suppression of $\kappa$ by an order of magnitude at room temperature. We obtain these results by perturbative calculations and non-perturbative Green’s function techniques on model spring-mass systems featuring tens of millions of atoms. Furthermore, we analytically determine the optimal correlations that minimize $\kappa$, and evaluate the impact of finite correlation lengths. Our quantitative theory can be generalized to force-constant disorder, providing a unifying framework to describe phonon-disorder scattering in materials featuring  nanostructures and complex nanoscale geometries. As such, it will allow theory to guide experiments towards a more refined control of lattice thermal transport. 

\textit{Formalism.} Going beyond the rule of thumb relating defect size with phonon wavelength, we will explain more precisely how spatial correlations affect phonon scattering. For simplicity, we consider in this work a generic spring-mass model of mass-disordered alloy on a simple cubic lattice with five different atomic masses in roughly equal proportions and other physical parameters comparable to SiGe alloys (see the supplementary material \cite{SM} for details). The presence of mass disorder couples two phonon modes of wavevectors $\bm{k}$ and $\bm{k+q}$ with a coupling matrix element $g_{\bm{q}}$, allowing mode $\bm{k}$ to scatter into mode $\bm{k+q}$. In second-order perturbation theory, the phonon-disorder transport scattering rate for mode $\bm{k}$ is given by $1/\tau^\text{d,tr}_{\bm{k}} = \frac{\pi}{\omega_{\bm{k}}} \sum_{\bm{q}} \big\langle \left| g_{\bm{q}} \right|^2 \big\rangle (1-\cos(\theta_{\bm{k}+\bm{q},\bm{k}})) \delta(\omega_{\bm{k}+{q}}^2 - \omega_{\bm{k}}^2)$ where $\theta_{\bm{k}+\bm{q},\bm{k}}$ is the angle between the velocities of modes $\bm{k}$ and $\bm{k+q}$. The factor $1-\cos(\theta_{\bm{k}+\bm{q},\bm{k}})$, often neglected, accounts for the fact that forward scatterings contribute less to thermal resistivity than backward scattering, and must be included here for reasons that are to become clear. In this expression, only the disorder-average of $\left|g_{\bm{q}}\right|^2$ matters:
\begin{equation}
\label{couplings}
\left\langle \left| g_{\bm{q}} \right|^2 \right\rangle = \omega^4 \frac{1}{N} \sum_{\bm{r}} C(\bm{r}) e^{-i \bm{q} \cdot \bm{r}}
\end{equation}
The right-hand side features the spatial Fourier transform (FT) of the mass correlation function $C(\bm{r}) = \left\langle \delta m_{\bm{r}} \, \delta m_{\bm{0}} \right\rangle$, where $\delta m_{\bm{r}}$ is the relative mass perturbation at site $\bm{r}$. In the absence of correlations, $C(\bm{r}) = \left\langle \delta m^2 \right\rangle \delta_{\bm{r} \bm{0}}$ and only the $\omega^4$ factor remains, leading to Rayleigh scattering and to weak scattering of heat-carrying low-frequency acoustic modes. However, it is possible to manipulate and redistribute the matrix elements within the first Brillouin zone by introducing correlations, i.e., going beyond randomly distributed defects. In particular, irrespective of the short-range behavior, the FT of a long-range power-law decaying correlation function $C(\bm{r}) \propto 1/r^\alpha$ behaves in three dimensions as $1/q^{3-\alpha}$, concentrating the matrix elements close to the Brillouin zone center if $\alpha < 3$. Thus, low-energy phonons are scattered more effectively at the price of increased forward scattering for the high-energy phonons. In this case, second-order perturbative calculations of the low-frequency phonon-disorder transport scattering rates suggest that they should decrease as $\omega^{1+\alpha}$ instead of $\omega^4$ \cite{SM}, consistent with previous perturbative calculations \cite{johnWavePropagationLocalization1983,chuEffectCorrelationsLocalization1989}. This modified power law implies much more efficient scattering of the heat-carrying acoustic phonons than from Rayleigh scattering, leading to reduced thermal conductivity. 

To test this idea, we built large disordered supercells spanning tens of millions of lattice sites in which we introduced mass disorder characterized by power-law decaying correlation functions. We also considered uncorrelated mass disorder as a reference. From the real-space dynamical matrix of these supercells, we used the Chebyshev polynomials Green's function method (CPGF) to obtain the phonon Green's function by a numerical expansion on the basis of Chebyshev polynomials \cite{ferreiraCriticalDelocalizationChiral2015,weisseKernelPolynomialMethod2006,bouzerarDrasticEffectsVacancies2020,thebaudPerturbationTheoryThermal2022,thebaudSuccessBreakdownTmatrix2020}. This approach is nonperturbative, allowing a full treatment of disorder effects. From the Green's function, we calculated the phonon spectral function, the phonon lifetimes, and the frequency-dependent thermal conductivity. We also obtained $\kappa$ from the Boltzmann transport equation (BTE) under the relaxation time approximation using transport scattering rates calculated from Fermi's golden rule (FGR). More details on the methodology, formalism and computational limitations can be found below and in~\cite{SM}. Of note, we use the term 'inverse lifetime' here rather than 'scattering rate' for phonon-disorder spectral linewidths to emphasize the distinction with 'transport scattering rate' that includes a vertex correction factor.  

\textit{Results and discussion.} The top panel of Fig.~\ref{fig2} shows two-dimensional cuts of the mass distribution in supercells with $200^3$ atoms characterized by a correlation function $C(\bm{r}) \propto 1/r^\alpha$ with $\alpha = \infty$ (uncorrelated) and $\alpha = 2$, $1$, and $0.5$. The mass distributions, involving five atomic species, are implemented 'by hand' to give the desired correlations (see~\cite{SM} for details). The mass distribution in the uncorrelated case is simply white noise, but as more and more long-range correlations are introduced, clusters and nanostructures of heavy and light atoms appear and grow larger, keeping however a broad size distribution. The bottom panel of Fig.~\ref{fig2} shows the corresponding CPGF-calculated phonon spectral function $A(\bm{k},\omega)$ on a high-symmetry path in the first Brillouin zone. For a given plane-wave mode of wavevector $\bm{k}$, $A(\bm{k},\omega)$ represents the energy distribution of that mode, and is close to the dynamical structure factor probed by inelastic X-ray and neutron scattering \cite{taraskinConnectionTrueVibrational1997,mutkaGeneralizedDensityofstatesAnharmonicity2008}. In the absence of disorder, plane waves are normal modes of the system so $A(\bm{k},\omega) = \delta(\omega - \omega_{\bm{k}})$. In the presence of disorder, the spectral peak is shifted (phonon frequencies are renormalized) and broadened (phonons acquire a finite lifetime). For visualization purposes, an artificial Lorentzian broadening of $\SI{1}{rad.THz}$ is added to the physical broadening from mass disorder in Fig.~\ref{fig2}. In all cases, low-frequency acoustic phonons are well defined in the sense that their broadening is much smaller than their central frequency, and they match the dispersion given by the virtual crystal approximation (VCA), which simply averages the atomic masses \cite{abelesLatticeThermalConductivity1963} and for our single-site model results in three degenerate acoustic phonon branches. The uncorrelated spectrum (leftmost panel) features ill-defined phonons above \SI{40}{rad.THz} whose broadening is comparable to their frequency (diffusons in Allen and Feldman's terminology \cite{allenDiffusonsLoconsPropagons1999}), and some flat branches above the VCA spectrum arising from isolated light atoms. The most obvious effect of long-range correlations is to introduce more structure to the flat branches above the VCA. This indicates that, for $1/\sqrt{r}$ correlations, the phonons almost start to propagate inside the domains as if they were infinite, leading to emergent light and heavy dispersions \cite{thebaudSuccessBreakdownTmatrix2020,kormannPhononBroadeningHigh2017}. Nevertheless, correlations do not seem to drastically alter the lifetimes of high-frequency VCA modes. 

\begin{figure}
\centering
\includegraphics[width=1.0\columnwidth]{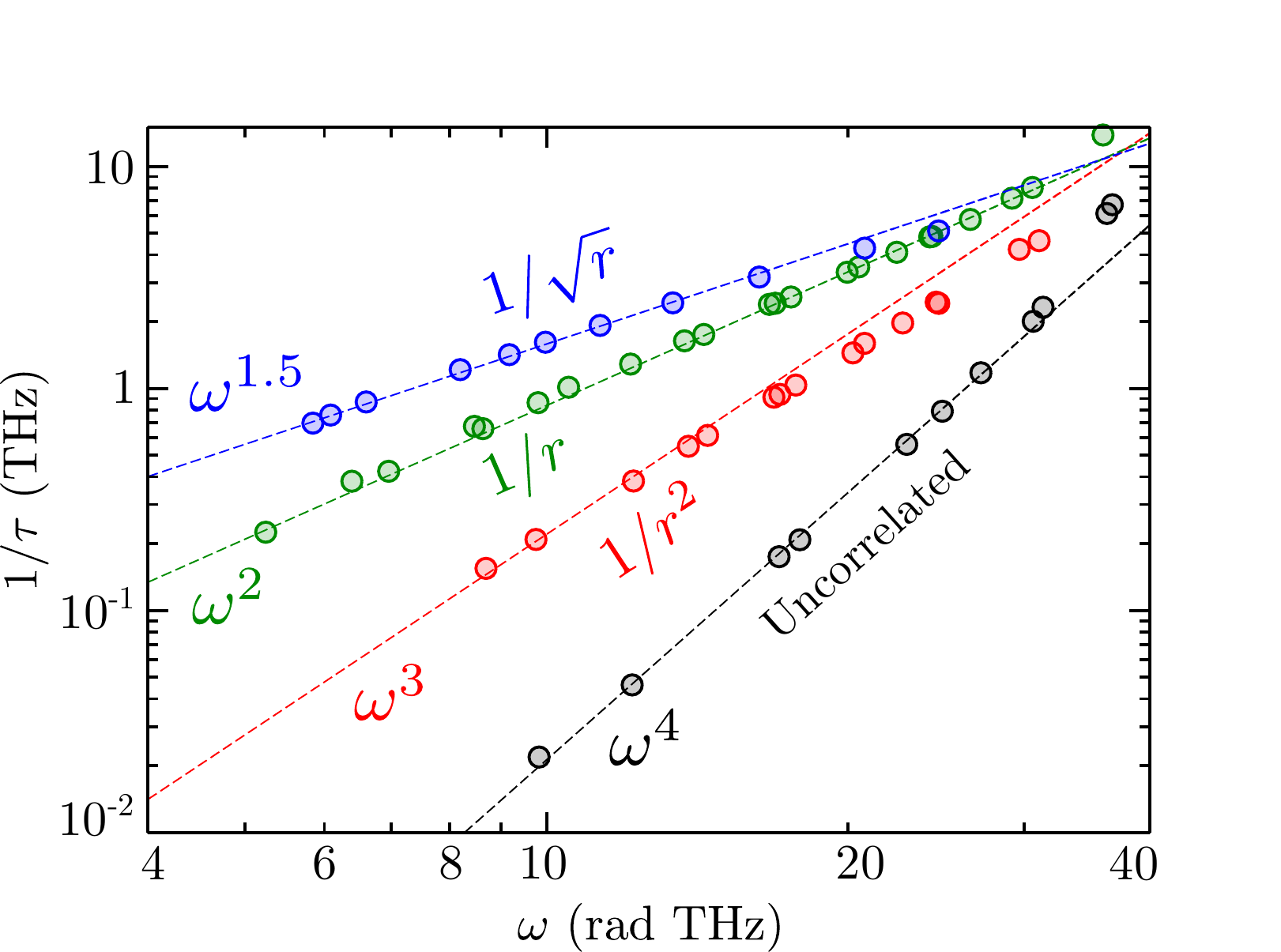}
\caption{Phonon-disorder inverse lifetimes extracted from the width of the CPGF-calculated spectral peaks for several power-law decaying correlations.}
\label{fig3}
\end{figure}

For phonon modes below \SI{40}{rad.THz} whose full width at half maximum (FWHM) can be defined, we extracted the FWHM that corresponds to the disorder-limited inverse lifetime of the mode, shown in Fig.~\ref{fig3} (the spectral functions of selected modes in the relevant frequency range are shown in~\cite{SM}). In the uncorrelated case, we find the usual $\omega^4$ low-frequency behavior. However, we find an $\omega^3$ law for $1/r^2$ correlations, an $\omega^2$ law for $1/r$ correlations and an $\omega^{1.5}$ law for $1/\sqrt{r}$ correlations, with lifetimes orders of magnitude smaller than in the uncorrelated case. This validates the perturbative result that $1/r^\alpha$ correlations lead to a $\omega^{1+\alpha}$ power law, and confirms that it is possible to circumvent Rayleigh scattering by introducing long-range spatial correlations. 

\begin{figure}
\centering
\includegraphics[width=1.0\columnwidth]{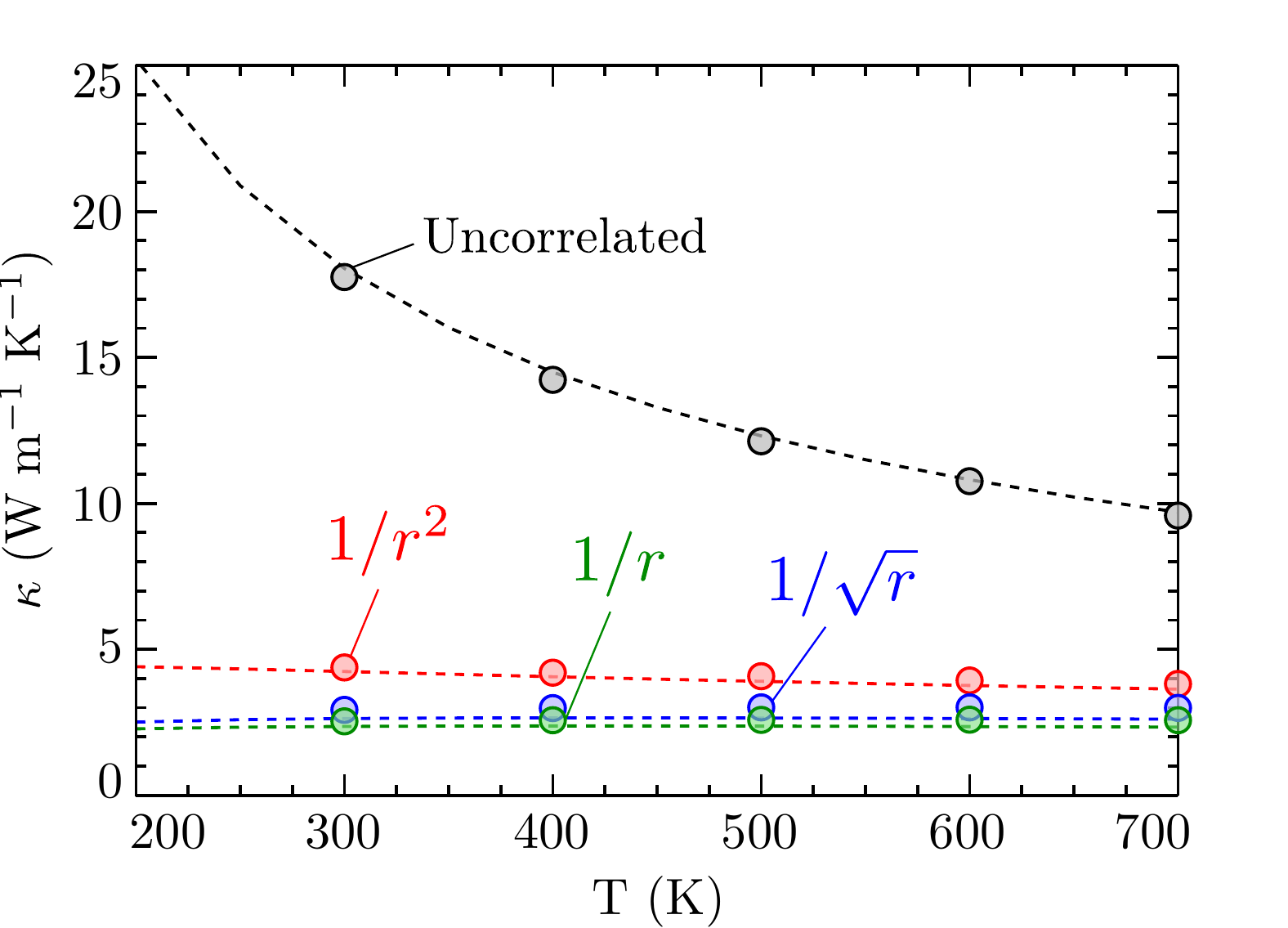}
\caption{The thermal conductivities for several power-law decaying correlations evaluated through Green-Kubo CPGF calculations above \SI{10}{rad.THz}, with second-order perturbation theory below \SI{10}{rad.THz} (dots) and through perturbation theory only (dashed lines).}
\label{fig4}
\end{figure}

From Fig.~\ref{fig3}, one could be tempted to conclude that the longer-range the correlations, the lower the thermal conductivity. As we shall see, however, this is not the case. Evaluating $\kappa$ in these systems is \textit{a priori} non-trivial, given the breakdown of the quasiparticle picture at high frequencies. To overcome this issue, we used the Green-Kubo formalism (quantum linear response theory), evaluating $\kappa$ directly by the CPGF method on supercells with $1200 \times 200 \times 200$ atoms \cite{SM}. This approach, which does not rely on the phonon quasiparticle picture, includes the diffusive channel (thermal transport by broadened overlapping modes \cite{simoncelliUnifiedTheoryThermal2019}) and is comparable to methods recently proposed by Isaeva \cite{isaevaModelingHeatTransport2019} and Caldarelli \cite{caldarelliManybodyGreenFunction2022a} (see Ref.~\cite{thebaudPerturbationTheoryThermal2022} for a discussion of the differences between these approaches). Crucially, it also takes into account all the so-called vertex corrections encoding the fact that forward scatterings contribute less to the thermal resistivity than backward scatterings \cite{ashcroftSolidStatePhysics1976}. Long-range correlations favor low-$\bm{q}$ matrix elements, leading to mostly forward scattering of large-wavevector phonons. Thus, neglecting the vertex corrections would introduce large errors in the evaluation of $\kappa$. At low-frequencies, this method agrees with the thermal conductivity obtained using the standard kinetic expression (or, equivalently, solving the Boltzmann transport equation under the relaxation time approximation) with phonon-disorder transport lifetimes evaluated through second-order perturbation theory \cite{thebaudPerturbationTheoryThermal2022}. Due to the presence of the $1-\cos(\theta)$ forward scattering, the most relevant vertex corrections (ladder diagrams) are also included in this simpler approach, and we use it for the contributions to $\kappa$ below \SI{10}{rad.THz} to avoid the prohibitive computational cost of the CPGF method at these frequencies. In all cases, we accounted for Umklapp anharmonic scattering through a simple expression $1/\tau_U = A \omega^2 T e^{-\Theta/T}$, with similar parameters as found in the literature for Si, Ge, and GaN compounds \cite{mingoCalculationSiNanowire2003,jeongFullDispersionDebye2011a,asen-palmerThermalConductivityGermanium1997,guthyUnusuallyLowThermal2008,luoGalliumArsenideThermal2013,gargRoleDisorderAnharmonicity2011a} (see also~\cite{SM}). The results are shown in Fig.~\ref{fig4}: dots correspond to the above method involving Green-Kubo CPGF calculations and the dashed lines to kinetic perturbative calculations only. Both methods are in very good agreement, indicating that the diffusive channel does not play an important role here and that the crucial physics is captured by the perturbative evaluation of the transport lifetimes. For this model, the room-temperature thermal conductivity is around \SI{18}{W.m^{-1}.K^{-1}} without correlations. However, by introducing $1/r$ correlations, it is reduced to \SI{2.5}{W.m^{-1}.K^{-1}}, almost an order of magnitude decrease. This suggests that spatially correlated point defects can indeed be used to crush the thermal conductivity. $1/r^2$ correlations lead to a large but somewhat less effective decrease, with \SI{4.4}{W.m^{-1}.K^{-1}}. Perhaps more surprisingly given the results of Fig.~\ref{fig3}, $1/\sqrt{r}$ correlations are also slightly less effective than $1/r$ correlations, with \SI{2.9}{W.m^{-1}.K^{-1}}. We show in~\cite{SM} the same graph in log scale making the difference between the long-range correlations clearer, with a brief discussion on the importance of vertex corrections.

Why are $1/\sqrt{r}$ correlations less effective despite the shorter acoustic phonon lifetimes displayed in Fig.~\ref{fig3}? The longer the correlation range, the more they concentrate the coupling matrix elements~\eqref{couplings} at the center of the Brillouin Zone (at small $\bm{q}$). This leads to more foward scattering of high-frequency phonons with large wavevectors, allowing these modes to contribute more efficiently to thermal transport. Thus, to minimize the thermal conductivity, correlations should strike a balance between scatterings at all wavelengths. More precisely, for a given disorder strength $\left\langle \delta m^2\right\rangle$, the Fourier transform $\tilde{C}({\bm{q}})$ of the mass correlation function $C({\bm{r}})$ must obey a constraint: $\frac{\Omega}{(2 \pi)^3} \int d^3 \bm{q} \, \tilde{C}({\bm{q}}) = \left\langle \delta m^2\right\rangle$, where $\Omega$ is the system size and the integral covers the Brillouin zone. If we assume that the thermal conductivity is given by the standard kinetic expression with perturbative transport phonon lifetimes (which yields excellent agreement with the full CPGF treatment, see Fig.~\ref{fig4}), finding the optimal correlation function to minimize $\kappa$ can then be recast as a classic optimization under constraint problem. We introduce a Lagrange multiplier $\lambda$ and set the functional derivative of $\kappa$ minus the constraint to $0$:
\begin{equation}
\label{kappa_min}
\frac{\delta}{\delta \tilde{C}({\bm{q}})} \left[\int \frac{d^3 \bm{k}}{(2 \pi)^3} \, c_v(\omega_{\bm{k}}) v_{\bm{k}}^2 \tau^\text{tr}_{\bm{k}}  - \lambda \int d^3 \bm{k} \, \tilde{C}({\bm{k}}) \right] = 0
\end{equation}
where $v_{\bm{k}}$ is the velocity of mode $\bm{k}$ along the transport direction and $c_v(\omega_{\bm{k}})$ the heat capacity of that mode. $\tau^\text{tr}_{\bm{k}} = (1/\tau_U + 1/\tau^\text{d,tr}_{\bm{k}})^{-1}$ is the transport lifetime of mode $\bm{k}$ with $1/\tau^\text{d,tr}_{\bm{k}} = \pi \omega_{\bm{k}}^3 \sum_{\bm{q}} \tilde{C}({\bm{q}}) (1-\cos(\theta_{\bm{k}+\bm{q},\bm{k}})) \delta(\omega_{\bm{k}+{q}}^2 - \omega_{\bm{k}}^2)$. Assuming now isotropic correlations and a Debye dispersion, eq.~\eqref{kappa_min} yields the condition $1/\tau^\text{tr}_{\bm{k}} \propto \sqrt{c_v(\omega_{\bm{k}})} \omega_{\bm{k}}^2$ (see~\cite{SM} for the derivation). For high enough temperatures, $c_v$ becomes weakly dependent on the frequency and the condition becomes that the total transport scattering is quadratic in frequency. If anharmonic scattering is also quadratic in $\omega$, as was supposed above, then the optimal correlations are those that lead to a phonon-disorder transport scattering rate proportional to $\omega^2$. In other words, this argument suggests that of all the possible types of isotropic correlations, $1/r$ power-law correlations are the most efficient at reducing the thermal conductivity. This is consistent with Fig.~\ref{fig4}, although due to the simplifying assumptions in the above argument, longer-range correlations may in practice be just as efficient. 

Although we have limited our investigation to the case of mass defects for simplicity, the language of spatial correlations is also applicable to force constant disorder, thus potentially encompassing all forms of static disorder. The question then arises: in which materials could favorable correlations appear or be engineered to lower $\kappa$? The Ising model is one theoretical context, widely applicable to many systems \cite{singhIsingModelBrief2020,zhangTheoryRelaxorferroelectricity2020}, whose mean-field solution just so happens to yield correlations of the form $\frac{1}{r} e^{-r/\xi}$ with $\xi$ the correlation length that diverges as $T$ is lowered towards the critical temperature $T_c$. We performed perturbative calculations showing that the presence of a correlation length is not a serious issue for reducing $\kappa$ as long as it is not very small (a few \si{\angstrom}, see~\cite{SM}). Therefore, if the system was quenched from a temperature just above $T_c$, $1/r$ correlations would be captured and the thermal conductivity would be crushed, provided the correlations are not destroyed by phenomena such as nucleation and spinodal decomposition \cite{zhengDefectEngineeringThermoelectric2021,androulakisSpinodalDecompositionNucleation2007,girardAnalysisPhaseSeparation2013}. Another situation where $1/r$ correlations arise is in the strain field around dislocations, leading to a scattering rate proportional to $\omega$ due to the one-dimensional nature of the defect \cite{hanusThermalTransportDefective2021}. This linear law might be considered very efficient at scattering low-frequency phonons, but in fact the translation symmetry along the dislocation line forbids many scattering processes, which leads to a lower scattering rate. Nevertheless, the presence of a dense network of non-parallel dislocations might restore some degree of isotropy while preserving long-range strain fields, potentially yielding low thermal conductivities \cite{basuImprovedThermoelectricPerformance2014}. So-called procrystals, characterized by local orientational sum rules, are the structural or chemical equivalent of pyrochlore spin-ice materials and feature $1/r^3$ correlations \cite{henleyCoulombPhaseFrustrated2010,simonovDesigningDisorderCrystalline2020,overyDesignCrystallikeAperiodic2016,keenCrystallographyCorrelatedDisorder2015,ormrodmorleyRingStructureSelected2020}. However, the presence of topological defects, whose density might be controlled by thermal treatments, could introduce longer-range correlations. Other situations potentially involving long-range and medium-range correlations are nanoparticles and extended defects \cite{kunduRoleLightHeavy2011a,guoMieScatteringPhonons2020,jengModelingThermalConductivity2008}, nanodomains in relaxor ferroelectrics and charge density wave materials \cite{cowleyRelaxingRelaxorsReview2011}, and stacks of twisted atomic monolayers \cite{kimExtremelyAnisotropicVan2021,chiritescuUltralowThermalConductivity2007,nguyenSynthesisPropertiesTurbostratically2010,hadlandUltralowThermalConductivity2019}. With this work, we provide a general, quantitative and unifying framework to understand the transport properties of disordered systems, and we open the door towards the systematic design of structure-induced transport functionalities.

\textit{Summary.} Using large-scale non-perturbative numerical simulations, we showed that the $\omega^4$ Rayleigh power law for the defect-limited inverse lifetimes of low-frequency acoustic phonons can be broken by long-range spatial correlations in the distribution of point defects. We proposed this as a strategy for tuning the lattice thermal conductivity of functional materials, and found an order of magnitude reduction of the conductivity through state-of-the art, fully quantum mechanical techniques. We showed analytically that correlations decaying as $1/r$ minimize the thermal conductivity by striking an optimal balance between scattering low-frequency and high frequency phonons. We posit that this framework can describe phonon conduction in a great variety of nanostructured systems and guide experimentalists towards a fine control of thermal transport in bulk materials.






\section*{ACKNOWLEDGEMENTS}

We thank German Samolyuk for helpful discussions. This research was supported by the U.S. Department of Energy, Office of Science, Basic Energy Sciences, Materials Sciences and Engineering Division. We used resources of the Compute and Data Environment for Science (CADES) at the Oak Ridge National Laboratory, which is supported by the Office of Science of the U.S. Department of Energy under Contract No. DE-AC05-00OR22725.

\begin{footnotesize}
\bibliographystyle{unsrt} 
\bibliography{main_manuscript.bbl}

\end{footnotesize}

\end{document}